\documentclass[12pt]{article}
\usepackage{amsmath}
\usepackage{amssymb}
\usepackage{epsfig}
\usepackage{euscript}
\usepackage{fancybox}
\usepackage{color}

\def\mathswitchr#1{\relax\ifmmode{\mathrm{#1}}\else$\mathrm{#1}$\fi}

\newcommand {\pslash}{\hbox{$\not\hbox{\kern-2.3pt $p$}$}}

\usepackage{cite}

\usepackage{epic}

\def\alf1{ {\alpha\over\pi} }

\begin{document}
\begin{titlepage}
\begin{flushright}
{\bf BU-HEPP-10-02 }\\
{\bf Apr., 2010}\\
\end{flushright}
 
\begin{center}
{\Large 
New Approach to GUTs $^{\dagger}$
}
\end{center}

\vspace{2mm}
\begin{center}
{\bf   B.F.L. Ward}\\
\vspace{2mm}
{\em Department of Physics,\\
 Baylor University, Waco, Texas, 76798-7316, USA}\\
\end{center}

\vspace{5mm}
\begin{center}
{\bf   Abstract}
\end{center}
We introduce a new string-inspired approach to the subject of grand unification
which allows the GUT scale to be small, $\lesssim 200$TeV, so that it is within
the reach of {\em conceivable} laboratory accelerated colliding beam devices. The key ingredient is a novel use of the heterotic string symmetry
group physics ideas to
render baryon number violating effects small enough to
have escaped detection to date. This part of the approach 
involves new unknown parameters to be tested experimentally. A possible hint at the existence of these new
parameters may already exist in the EW precision data comparisons
with the SM expectations.
\vspace{10mm}
\vspace{10mm}
\renewcommand{\baselinestretch}{0.1}
\footnoterule
\noindent
{\footnotesize
\begin{itemize}
\item[${\dagger}$]
Work partly supported 
by NATO Grant PST.CLG.980342.
\end{itemize}
}

\end{titlepage}

\def\Kmax{K_{\rm max}}\def\ieps{{i\epsilon}}\def\rQCD{{\rm QCD}}
\renewcommand{\theequation}{\arabic{equation}}
\font\fortssbx=cmssbx10 scaled \magstep2
\renewcommand\thepage{}
\parskip.1truein\parindent=20pt\pagenumbering{arabic}\par
\indent The success and structure 
of the Standard Model(SM)~\cite{sm1,qcd} suggest that
all forces associated with the gauge interactions therein may
be unified into a single gauge principle associated
with a larger group ${\cal G}$ which contains $SU(2)_{L}\times U(1)_Y\times SU(3)^c$
as a subgroup, where we use a standard notation for the SM gauge group.
This idea was originally introduced in the
modern context in Refs.~\cite{pati-salam,geor-glash} and continues to
be a fashionable area of investigation today, where 
approaches which unify the SM gauge forces with that of
quantum gravity are now in very much vogue via the superstring
theory~\cite{gsw,jp} and its various low energy 
reductions and
morphisms~\cite{jp}. In this paper, we focus only on the unification
of the SM gauge forces themselves, candidates for 
which we call as usual GUTs, 
leaving aside 
any possible unification with quantum gravity until a later
study~\cite{elswh}. 
\par
We need to admit at the outset that a part of our motivation is the recent
progress in approaches to the Einstein-Hilbert theory for quantum gravity
in which improved treatments of perturbation theory via resummation methods,
the asymptotic safety approach~\cite{asympsfty}, the resummed quantum
gravity approach~\cite{bw1} or the Hopf-algebraic Dyson-Schwinger
equation renormalization theory approach~\cite{kreimer}, 
and the introduction of
an underlying loop-space at Planck scales, loop quantum gravity~\cite{lpqg},
have shown that the apparently bad unrenormalizable
behavior of that theory may be cured by the dynamical interactions or
modifications 
within the
theory itself, as first anticipated by Weinberg~\cite{asympsfty}. 
Such progress would suggest that the unification of all other forces 
can be a separate problem
from the
problem of treating the apparently
bad UV behavior of quantum gravity. We explore this
suggestion in what follows.\par 
Our idea is to try to formulate GUTs so that they are accessible
to very high energy colliding beam devices such as the VLHC,
which has been discussed elsewhere~\cite{vlhc} with cms energies
in the 100-200TeV regime. We will show that we can achieve such
GUTs that satisfy the usual requirements: no anomalies,
unified SM couplings, baryon stability, absence/suppression 
of other unwanted transitions and naturalness requirements (this may
just mean N=1 susy here~\cite{wittn}). Here, we add the new condition
that the theory will live in 4-dimensional Minkowski space.
We call this our {\em known physical reality condition}. 
The most demanding requirement will be
seen to be baryon stability.\par

Indeed, let us just illustrate why the most difficult aspect of a 
GUT with a (several) hundred TeV
unification scale is the issue of baryon number stability: the proton
must be stable to $\sim 10^{29-33}$yrs, depending on the mode,
whereas the natural lifetime
for physics with a 100TeV scale for a dimension 6 transition 
in a state with the size and mass of the proton is
$\sim 0.01$yr for example. Evidently, some new mechanism is needed
to suppress the proton decay process here.\par

Rather than to 
move the GUT scale to $\sim 10^{13}$TeV as is usually done~\cite{gut1},
or invoke hitherto unknown phenomena, such as extra dimensions~\cite{kdgut,gut1}, extra vector representations of the gauge 
group~\cite{fr-min-all}
, etc., we will try to rely on well-tested ideas used in a novel way -
we will use what is sometimes called a radically conservative approach.
We look at the fundamental structure of a GUT theory. We notice that
it is organized by gauge sector, by family sector and by Higgs
sector for spontaneous symmetry breaking. Let us look at the
family and gauge sectors.\par

In Ref.~\cite{geor-glash}, the ${\bf 10+\bar{5}}$ of 
$SU(5)$ was advocated and shown
to accommodate the SM family with a massless neutrino. Recently, with
advent of neutrino masses~\cite{neut1}, we need to extend this to
a sixteen dimensional representation. We will use the ${\bf 16}$ of $SO(10)$~\cite{gross},
as it decomposes as  ${\bf 10+\bar{5}\bf+1}$ under an inclusion of $SU(5)$
into $SO(10)$. We know from the heterotic 
string formalism~\cite{gsw,jp}(we view here modern string theory 
as an extension of quantum field theory which can be used to abstract
dynamical relationships which would hold in the real world even if
the string theory itself is in detail only an approximate, mathematical 
treatment of that reality, just as the old strong interaction string theory~\cite{schwz1} could be used to abstract properties
of QCD~\cite{qcd} such as Regge trajectories even before QCD was discovered)
that in the only known and accepted unification of the SM and gravity,
the gauge group $E_8\times E_8$ is singled-out
when all known dualities~\cite{jp} are taken into account to relate equivalent superstring theories. A standard breakdown of this
symmetry to the SM gauge group and family structure is as follows~\cite{jp}:
\begin{equation}
\begin{split}
E_8&\rightarrow SU(3)\times E_6 \rightarrow SU(3)\times SO(10)\times U'(1)\\ 
&\rightarrow SU(3)\times SU(5)\times U''(1)\times U'(1)\\
&\rightarrow  SU(3)\times SU(3)^c\times SU(2)_L\times U(1)_Y\times U''(1)\times U'(1)
\end{split}
\label{brk1}
\end{equation}
where the SM gauge group is now called out as $SU(3)^c\times SU(2)_L\times U(1)_Y$. It can be shown that the ${\bf 248}$ of $E_8$ then splits
under this breaking into $({\bf 8,1})+({\bf 1,78})+({\bf 3,27})+(\bar{\bf{3}},
\overline{\bf{27}})$
under $SU(3)\times E_6$ and that each ${\bf{27}}$ under $E_6$ contains exactly
one SM family 16-plet with 11 other states that are paired with their anti-particles in helicity via real representations 
so that they would be expected to become massive at the GUT scale.
Let us consider that we have succeeded with the heterotic string breaking
scenario to get 3 families~\cite{gut1} 
under the first $E_8$ factor in the $E_8\times E_8$
gauge group. They are singlets under the second $E_8$.
We now repeat the same pattern for the second factor as well.
This gives us 6 families, one set of ``family 
triplets'' transforming non-trivially only under $E_{8a}$
and the other set of ``family triplets'' transforming non-trivially only
under $E_{8b}$,
where $E_8\times E_8\equiv E_{8a}\times E_{8b}$. To stop baryon instability,
we identify the light quarks as those from $E_{8a}$ and the light leptons
as those from $E_{8b}$ -- here, light means light on the scale of $M_{GUT}$,
the grand unified theory (GUT) scale. The remaining particles in each sector
are then at the respective
scales $M_{LM}$ between their current experimental limits and the
GUT scale. The proton can not decay because the leptons
to which it could transform via (leptoquark) bosons are all at too
high a scale. 
\par
\begin{figure}
\begin{center}
\epsfig{file=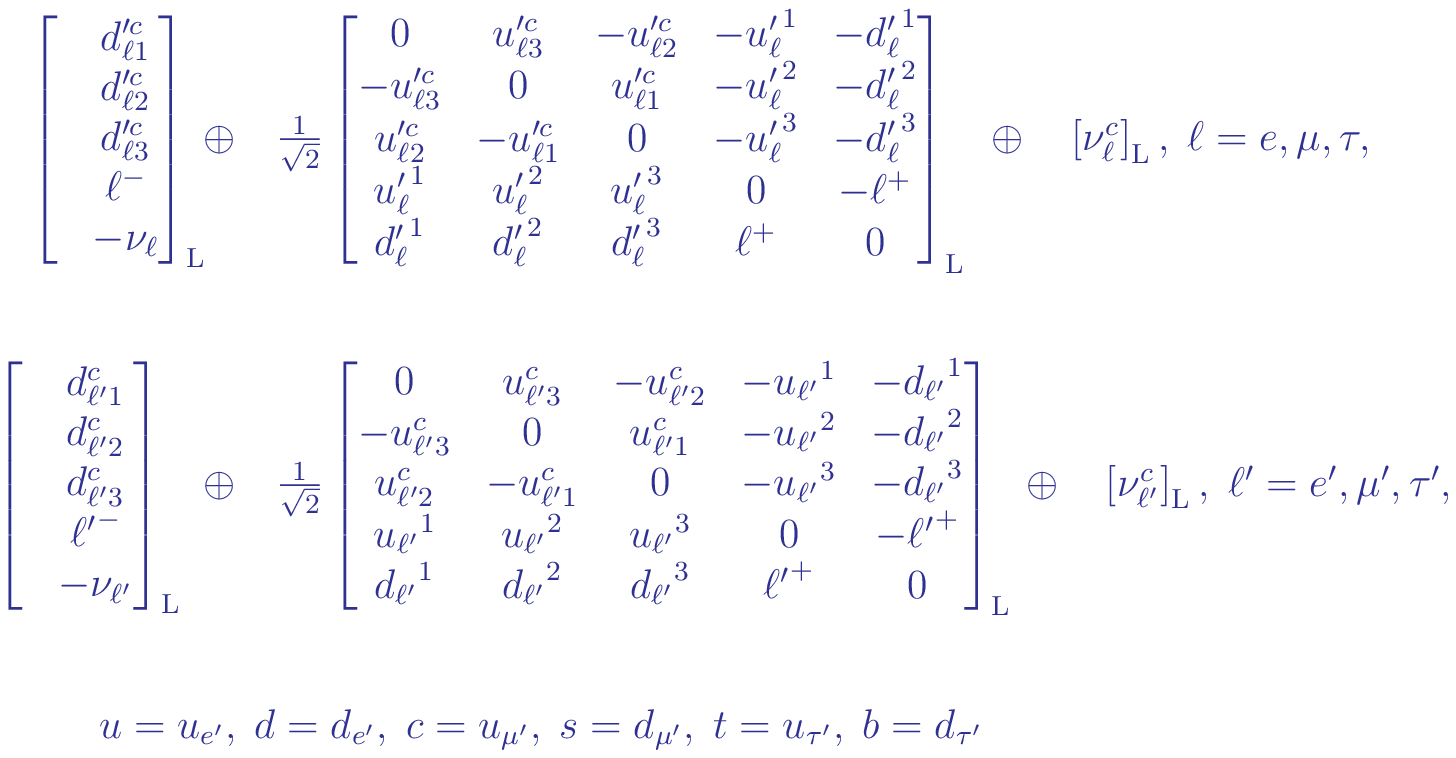,width=140mm}
\end{center}
\caption{\baselineskip=7mm     $SU(5)$ decomposition of 
six SM families with the prediction of 
six new heavy leptons and six new heavy quarks, all possibly 
accessible at the next
set of high energy colliders. The superscript $c$ denotes charge conjugation as usual and the color index runs from $1$ to $3$ here. 
When the effective high energy gauge symmetry is
$E_8\times E_8\times E_8\times E_8 \equiv E_{8,1}\times E_{8,2}\times E_{8,3}\times E_{8,4}$, the structure in this figure occurs twice, but with new quarks and leptons
as yet unseen.}
\label{fig1}
\end{figure}

Already, let us note that, while our approach to proton stability is very much related to the approaches in Refs.~\cite{fr-min-all}
, it differs from the standpoint of radical conservatism - we only use the family structure that has been seen in Nature in the
standard SM families: $\{{\bf 10+\bar{5}\bf+1}\}_j$, where we now have six of them instead of the usual three so that the family index $j$ now runs from $j=1$ to $j=6$
and we do not have any vector representations, as we 
illustrate in Fig.~\ref{fig1}. For this reason we have no 
problem with such issues as charge quantization. We are predicting 
the discovery of the {\it equivalent} of 3 new families of quarks 
and leptons at the next very high energy machines with gauge quantum 
numbers entirely the same as those that have already been seen 
in Nature but with significantly larger masses.
In particular, we note that just a single Higgs $SU(2)_L$
doublet, as originally proposed
by Weinberg and Salam~\cite{gsw}, 
is enough to give all particles their masses in a large region of our parameter space. Of course,
this does not exclude more than one such doublet.
\par
It is also important to stress that we are only abstracting the family and gauge structure of the breaking pattern in (\ref{brk1})(we discard everything else), very much in the spirit of Gell-Mann's abstraction of the algebra of currents from free field theory in Ref.~\cite{gellmn} for the strong interaction, without claiming that the details
in the breaking are themselves also relevant. Indeed, our entire point is that these details, whether they are from the string theoretic perspective or the usual GUT perspective, may not be relevant at all.\par
Since we are entering into some 
discussion about hitherto unexplored phenomena,
we have to be open about the framework in the most basic ways.
For example, the heterotic superstring Lagrangian has the 
action, using the language of conformal
field theory for definiteness~\cite{jp}, for the matter sector (as opposed to the ghost sector) given by 
\begin{equation}
S(X,\tilde{\psi},\lambda)=\frac{1}{4\pi}\int d^2z\left(\frac{2}{\alpha'}\partial X^\mu\bar{\partial}X_\mu+\lambda^A\bar{\partial}\lambda^A+\tilde{\psi}^\mu{\partial}\tilde{\psi}_\mu\right)
\label{htric}
\end{equation}
where we have introduced the fields $X^{\mu}(z,\bar{z}),\tilde{\psi}^\mu(\bar{z}),\; \mu=0,\ldots,9$ for the left-moving part of the bosonic string and the
right-moving part of the type II superstring, respectively, as well as the 32 left-moving
spin-$\frac{1}{2}$ fields $\lambda^A,\; A=1,\ldots,32$, with the
boundary conditions $\lambda^A(w+2\pi)=\eta\lambda^A(w),\; A=1,\ldots,16,\;
\lambda^A(w+2\pi)=\eta'\lambda^A(w),\; A=17,\ldots,32,$
where $\eta$ and $\eta'$ each are $\pm 1$, $e^{-iw}=z$ and $\frac{1}{2\pi\alpha'}$ is the usual string tension.
The $\lambda^A$ are needed 
to complete the cancellation of central charge when all ghosts are taken into account and the boundary conditions, with the 
attendant GSO projections, just give us the $E_8\times E_8$ heterotic
superstring theory~\cite{gsw} as it is well known~\cite{jp}. 
Here, we extend this to the possibility that we have two
such contributions to the world action, two strings, 
that for the moment will
be non-interacting copies of each other:$S_{\text{world}}=S(X(1),\tilde{\psi}(1),\lambda(1))+S(X(2),\tilde{\psi}(2),\lambda(2))$
where each $\{X(j),\tilde{\psi}(j),\lambda(j)\}$ is an independent 
copy of the heterotic string fields in
(\ref{htric}).
The gauge group of the world 
is then two copies of $E_8\times E_8$\footnote{If one wants to avoid any reference to superstring theory, one can just postulate our symmetry and families as needed, obviously, with the effective GUT gauge group $SO(10)\times SO(10)\times SO(10)\times SO(10)$ with discrete symmetery used to achieve equality of
the gauge couplings at the GUT scale and the textbook~\cite{gross} 
symmetry breaking
to the respective SM gauge group factors ; we leave this to the discretion of the reader.}. If we repeat the
model construction above, we have the possibility
of making 6 light families, three of which we have not seen,
so we take it that they may be at any scale above what
has been eliminated up to the GUT scale. They may appear at LHC, for example.
\par
The ordinary electroweak and strong interaction
gauge bosons are now an unknown mixture
of the two copies of two sets of such bosons from the two $E_8's$
associated to a given string Lagrangian:
when we break the four $E_8$'s  each to a product group
$SU(3)\times E_6$ and then subsequently break each of
the four $E_6$'s to get four copies 
of ${SU(3)}^c\times SU(2)_L\times U(1)_Y$, for
the initially massless gauge bosons for $SU(3)^c_i\times SU(2)_{Li}\times U(1)_{Yi}\in E_{8,i}$,~$G^a_i,~a=1,\cdots,8,~A^{i'}_i,~i'=1,\cdots,3,~B_i$,\ $i=1,\cdots,4$, in a standard notation, we assume a further breaking at the 
GUT scale so that the following linear combinations are massless at the GUT scale $M_{GUT}$ while the orthogonal linear combinations acquire masses ${\cal O}(M_{GUT})$ --
\begin{equation}
\label{fgauge1}
\begin{split}
A_f^{i'}=\sum_{i=1}^{4}\eta_{2i} A^{i'}_i,\; B_f= \sum_{i=1}^{4}\eta_{1i} B_i .
\end{split}
\end{equation}
The mixing coefficients $\{\eta_{aj}\}$ satisfy
$\sum_{i=1}^{4}\eta_{ai}^2=1,\; a=1,2$.
By the discrete symmetry that obtains if the strings are identical copies of each other, all color and electroweak gauge couplings at the scale $M_{GUT}$ 
satisfy the usual GUT relations as first given by Georgi and Glashow in Ref.~\cite{geor-glash}.
\par
For the strong interaction, we take the minimal view that 
the quarks in each set of three families from the four $E_8$'s are confined.
By our discrete symmetry all four strong interaction
gauge couplings to be equal at the GUT scale. This
means that for the known quarks, we have gluons $G^a_1$. Of course,
experiments may ultimately force us to break the as yet unseen
color groups. This is straightforward to do following Ref.~\cite{LFLi}.\par
For the low energy EW bosons, we have quite a bit of 
freedom in (\ref{fgauge1}). We note the following values~\cite{siggi,pdg08} 
of the known
gauge couplings at scale $M_Z$: 
\begin{equation}
\label{fgauge2}
\begin{split}
&\alpha_s(M_Z)|_{\overline{MS}}= 0.1184\pm 0.0007\\
&\alpha_W(M_Z)|_{\overline{MS}}= 0.033812\pm 0.000021\\
&\alpha_{EM}(M_Z)|_{\overline{MS}}= 0.00781708\pm0.00000098
\end{split}
\end{equation}
It is well-known~\cite{gqw} that the factor of almost 4 between $\alpha_s(M_Z)$ and $\alpha_W(M_Z)$ and between $\alpha_W(M_Z)$ and $\alpha_{EM}(M_Z)$ when the respective unified values
are 1 and 2.67 require $M_{GUT}\sim 10^{13}-10^{12}$TeV. Here, with the use of the $\{\eta_{kj}\}$ we can absorb most of the discrepancy between the unification and
observed values of the coupling ratios so that the GUT scale is not beyond
current technology for accelerated colliding beam devices.\par
More precisely, we can set 
\begin{equation}
\label{fgauge3}
\begin{split}
\eta_{21}=\eta_{22}\cong \frac{1}{\sqrt{2.032}}\\
\eta_{11}=\eta_{12}\cong  \frac{1}{\sqrt{3.341}}
\end{split}
\end{equation} 
and this will leave a ``small'' amount of evolution do be done
between the scale $M_Z$ and $M_{GUT}$.\par
Indeed, with the choices in (\ref{fgauge3}), and the use of the one-loop 
beta functions~\cite{qcd}, if we use continuity of the gauge coupling constants at mass thresholds with one such threshold at $m_H\cong 120$GeV and a second one at $m_t=171.2$GeV for definiteness to illustrate our approach\footnote{Here, we take the limit that $M_{LM}$ is near $M_{GUT}$ for the illustration -- the case where it is a few TeV is done in the first Note-Added for completeness}, then the GUT scale can be easily evaluated to be $M_{GUT}\cong 100$TeV, as advertised. For, we get,
\begin{equation}
\label{fgauge4}
b^{U(1)_Y}_{0} = \frac{1}{12\pi^2}\begin{cases}4.385&,\;M_Z\le \mu\le m_H\cong 120\text{GeV}\\ 4.417&, \; m_H< \mu \le m_t\\ 
5.125&, \; m_t < \mu \le M_{\text{GUT}}
\end{cases}
\end{equation}
from the standard formula~\cite{qcd} 
\begin{equation}
\label{fgauge5}
b^{U(1)_Y}_{0}=\frac{1}{12\pi^2}\left(\sum_j n_j\left(\frac{Y_j}{2}\right)^2\right)
\end{equation}
where $b^{U(1)_Y}_0$ is the coefficient of $g'^3$ in the beta function for the
$U(1)_Y$ coupling constant 
$g'$ in the $SU(2)_{L}\times U(1)_Y$ EW theory of 
Glashow, Salam and Weinberg~\cite{sm1}, 
$n_j$ is the effective number of Dirac fermion degrees of freedom,
i.e., a left-handed Dirac fermion counts as $\frac{1}{2}$, a complex scalar
counts as $\frac{1}{4}$, and so on. Similarly, for the QCD and $SU(2)_{L}$
theories, we get the analogous
\begin{equation}
\label{fgauge6a}
b^{SU(2)_{L}}_{0} = \frac{-1}{16\pi^2}\begin{cases}3.708&,\;M_Z\le \mu\le m_H\cong 120\text{GeV}\\ 3.667&, \; m_H< \mu \le m_t\\ 
3.167&, \; m_t < \mu \le M_{\text{GUT}}
\end{cases}
\end{equation}
\begin{equation}
\label{fgauge6b}
b^{QCD}_{0} = \frac{-1}{16\pi^2}\begin{cases}7.667&,\;M_Z\le \mu\le m_t\\ 7&, \; m_t < \mu \le M_{\text{GUT}}
\end{cases}
\end{equation}
from the standard formula~\cite{qcd} 
\begin{equation}
\label{fgauge7}
b^{\cal H}_{0}=\frac{-1}{16\pi^2}\left(\frac{11}{3}C_2({\cal H})-\frac{4}{3}\sum_j n_jT(R_j)\right)
\end{equation}
where $T(R_j)$ sets the normalization of the generators $\{\tau^{R_j}_a\}$ of 
the group ${\cal H}$ in the representation $R_j$ via $\text{tr}\tau^{R_j}_a\tau^{R_j}_b=T(R_j)\delta_{ab}$ where $\delta_{ab}$ is the Kronecker delta and
$C_2({\cal H})$ is the quadratic Casimir invariant eigenvalue for the adjoined representation of ${\cal H}$. 
These results 
(\ref{fgauge4},\ref{fgauge5},\ref{fgauge6a},\ref{fgauge6b},\ref{fgauge7})
together with the standard one-loop solution~\cite{qcd}
\begin{equation}
\label{fgauge8}
g^2_{\cal H}(\mu)=\frac{g^2_{\cal H}(\mu_0)}{1-2b^{\cal H}_0 g^2_{\cal H}(\mu_0)\ln(\mu/\mu_0)}
\end{equation} 
allow us to compute the value $M_{GUT}\cong 100$TeV for the values of 
$\eta_{aj}$ given in (\ref{fgauge3}). Here, we use standard notation
that $g^2_{\cal H}(\mu)$ is the squared running coupling constant at scale $\mu$
for ${\cal H}=U(1)_Y, \; SU(2)_{L}, \; QCD\equiv SU(3)^c$ and
we note as well that
the parameters $\eta_{aj}$ modify the usual unification conditions
at the GUT scale, for $\eta_{a1}=\eta_{a2},\; a=1,2$, via
\begin{equation}
\begin{split}
\alpha_{QCD}(M_{GUT})=\frac{1}{\eta_{21}^2}\alpha_{SU(2)_L}(M_{GUT}),\\
\alpha_{QCD}(M_{GUT})=\frac{5}{3\eta_{11}^2}\alpha_{U(1)_{Y}}(M_{GUT}),
\end{split}
\end{equation}
where as usual $\alpha_{\cal H}(\mu)\equiv g^2_{\cal H}(\mu)/(4\pi)$.
The remaining parameters $\eta_{aj},\;a=1,2,\; j=3,4$ are such that
the conditions $\sum_j\eta_{aj}^2=1, \;a=1,2$ hold and would be subject to
investigations of the higher energy multiplets/massive gauge bosons 
that have yet to be discovered according to the model we present here. We note the value $\alpha_{QCD}(M_{GUT})= 0.0613$ for the case presented here, for reference. Its dependence on the $\eta_{aj}$
can be seen, in the current example, from the result
\begin{equation}
\alpha_{QCD}(M_{GUT})=\frac{\alpha_{QCD}(m_t)}{1+\frac{b_0^{QCD}}{b_0^{QCD}-\eta_{21}^2b_0^{SU(2)_L}}\left(\frac{\eta_{21}^2\alpha_{QCD}(m_t)}{\alpha_{SU(2)_L}(m_t)}-1\right)},
\end{equation}
from which one can see why $M_{GUT}$ is significantly lowered
by the values of $\eta_{aj}$ that we use. As usual, $\alpha_s\equiv \alpha_{QCD},\; \alpha_W=\alpha_{SU(2)_L}$.
\par
We note that the value of $100$TeV for the unification scale has been chosen
for illustration, as in principle any value between the TeV scale and 
the Planck scale is allowed in our approach. Experiment would tell us what the true value is.\par
In principle the problem with baryon stability could re-appear if
the leptoquark bosons from different $E_{8,i}$ would mix. To prevent this, it is enough that the B-L charge from each $E_{8,i}$ is separately conserved,
so that leptoquarks from different $E_{8,i}$ cannot mix.
\par
We sum up with an interesting possible application of our approach. 
We recall the very precise values of the EW parameter $\sin^2\theta^{\text{lept}}_{W,\text{eff}}$ 
from the lepton sector via the $A_{LR}$ and 
from the precision hadronic observable $A^b_{FB}$ as
summarized in Ref.~\cite{lewwg}. 
These two measurements, arguably two of the most precise measurements 
at SLC and LEP, disagree by 3.2 $\sigma$,
where the two respective values are~\cite{lewwg} 
$0.23098\pm 0.00026$
and $0.23221\pm 0.00029$. We see above that just a small change in the 
mixing coefficients for gauge bosons attendant to 
the families with the light quarks versus those for gauge 
bosons attendant to the families with light leptons easily 
accommodates any actual difference in these two measurements. 
A more precise set
of EW measurements, such as those possible at an ILC/CLIC
high energy $e^+e^-$ colliding beam device, 
would eventually clarify the situation, presumably.
More importantly, we propose here a ``green pasture'' instead of the traditional ``desert''~\cite{geor-glash,gqw}.\par
\section*{Notes Added}
\begin{itemize}
\item In the text, we took the intermediate scale $M_{LM}$
to the limiting value $M_{GUT}$ so that the new quarks and leptons 
we predict are all at
the GUT scale and do not enter into our running coupling constant analyses
for simplicity to illustrate the basic ideas of the discussion. It is straightforward to redo the analyses to allow the more interesting case where for example
we put one set of new leptons and new quarks at $M_{LM}=2$TeV, so that they would be accessible at the LHC. Then by evolving our coupling constants first from $\mu=m_t$ to $\mu=2$TeV using the results given in the text and then from $\mu=2$TeV to $\mu=M_{GUT}=100$TeV with the new values $(b^{U(1)_Y}_{0},b^{SU(2)_{L}}_{0},b^{QCD}_{0})=(10.1875/(12\pi^2),(5/6)/(16\pi^2),(-3/(16\pi^2)))$, we get that the required 
values of the $\eta_{ij}$
change to $\eta_{21}=\eta_{22}\cong \frac{1}{\sqrt{2.218}}$ and
$\eta_{11}=\eta_{12}\cong  \frac{1}{\sqrt{3.760}}$. LHC may therefore very well
discover some of our new states.
\item It is possible that the 3.2$\sigma$ effect discussed above is just a 
statistical fluctuation. Without this effect, we can simplify our approach as follows. We take a six light family string 
compactification~\cite{jp,gut1} in the first
$E_{8a}$ breaking and we leave open what number of light families we get from the breaking of the second $E_{8b}$ factor for a single heterotic string.
We then only have two sets of SM gauge bosons at $M_{GUT}$. Again, with the 3 families with the known light quarks we associate heavy leptons at scale $M_{LL}$ and with the three families with the known light leptons we associate heavy quarks at the scale $M_{QL}$
where in the text we set generically $M_{LL}\sim M_{QL}\sim M_{LM}$. The mixing formulas (\ref{fgauge1}) now just involve $\{\eta_{jk}, j=1,2,k=1,2\}$ 
and the known light quarks and leptons have the 'same' 
leptonic effective weak mixing angle. The same calculations
as we presented above in the text 
still obtain: for illustration with $ M_{LM}\sim M_{GUT}$, if we take now
$\eta_{21}=1/\sqrt{2.000},\eta_{11}=1/\sqrt{3.260}$,
we get $M_{GUT}\cong 136$TeV which is again in the $100-200$TeV regime. Our 'broken family' hypothesis again
realizes a 'green pasture' instead of the traditional 'desert'.
\end{itemize}


\begin{thebibliography}{99}
\bibitem{sm1}S.L. Glashow, Nucl. Phys. {\bf 22} (1961) 579; 
S. Weinberg, Phys. Rev. Lett. {\bf 19} (1967) 1264;
A. Salam, in {\em Elementary Particle Theory}, ed. N. Svartholm
(Almqvist and Wiksells, Stockholm, 1968), p. 367;
G.~'t Hooft and M.~Veltman, Nucl. Phys. B{\bf 44},189 (1972)
and {\bf 50}, 318 (1972); 
G.~'t Hooft, {\it ibid.} {\bf 35}, 167 (1971); M.~Veltman, {\it ibid.} {\bf 7}, 637 (1968). 
\bibitem{qcd}
D. J. Gross and F. Wilczek, 
Phys. Rev. Lett. {\bf 30} (1973) 1343;
H. David Politzer, {\it ibid.}{\bf 30} (1973) 1346; see also
, for example, F. Wilczek, in {\em Proc. 16th International Symposium on Lepton and 
Photon Interactions, Ithaca, 1993}, eds. P. Drell and D.L. Rubin 
(AIP, NY, 1994) p. 593, and references therein.
\bibitem{pati-salam} J. C. Pati and Adbus Salam, Phys. Rev. D{\bf 8}, 1240 (1973).
\bibitem{geor-glash} H. Georgi and S. L. Glashow, Phys. Rev. Lett. {\bf 32}, 438 (1974).
\bibitem{gsw} M.B. Green and J. H. Schwarz, Phys. Lett. B{\bf 149}, 117 (1984); {\it ibid.} {\bf 151}, 21 (1985); D.J. Gross {\em et al.}, Phys. Rev. Lett. {\bf 54}, 502 (1985); Nucl. Phys. B{\bf 256}, 253 (1985); {\it ibid.} {\bf 267}, 75 (1986); see also M. Green, J. Schwarz and E. Witten,
{\em Superstring Theory, v. 1 and v.2}, ( Cambridge Univ. 
Press, Cambridge, 1987 ) and references therein.
\bibitem{jp}
See, for example, J. Polchinski, {\em String Theory, v. 1 and v. 2},
(Cambridge Univ. Press, Cambridge, 1998), and references therein.
\bibitem{elswh} B.F.L. Ward, to appear.
\bibitem{asympsfty} S. Weinberg, in {\it General Relativity}, eds. S.W. Hawking
and W. Israel,(Cambridge Univ. Press, Cambridge, 1979) p.790; A. Bonanno and M. Reuter, Phys. Rev. D{\bf 65} (2002) 043508; J. Phys. Conf. Ser. {\bf 140} 
(2008) 012008; Phys. Rev. D{\bf 62} (2000) 043008; M. Reuter, Phys. Rev. D{\bf 57} (1998) 971; 
O. Lauscher and M. Reuter, {\it ibid.} {\bf 66} (2002) 025026, and references
therein; D. F. Litim, Phys. Rev. Lett.{\bf 92}(2004) 201301; Phys. Rev. D{\bf 64} (2001) 105007 and references therein; R. Percacci and D. Perini, Phys. Rev. D{\bf 68} (2003) 044018; A. Codello, R. Percacci and C. Rahmede, Ann. Phys. {\bf 324} (2009) 414; P. F. Machado and R. Percacci, Phys. Rev. D{\bf 80} (2009) 024020; R. Percacci, arXiv:0910.4951; G. Narain and R. Percacci, Class. Quant. Grav. {\bf 27} (2010) 075001, and references therein; J. Ambjorn, J. Jurkiewicz and R. Loll, arXiv:1004.0352, and referenecs therein. 
\bibitem{bw1} B.F.L. Ward, Mod. Phys. Lett. A{\bf 17} (2002) 2371;
Open Nucl. Part. Phys. J {\bf 2} (2009) 1; J. Cos. Astropart. Phys.{\bf 0402} (2004) 011; Mod. Phys. Lett. A{\bf 23} (2008) 3299, and references therein. 
\bibitem{kreimer} D. Kreimer, Ann. Phys. {\bf 323} (2008) 49; {\it ibid.} {\bf 321} (2006) 2757.
\bibitem{lpqg} T. Thiemann, in {\it Proc. 14th International Congress on Mathematical Physics}, ed. J.-C. Zambrini,(World Scientific Publ. Co., Hackensack, 2005) pp. 569-83; L. Smolin, hep-th/0303185; A. Ashtekar and J. Lewandowski, Class. Quantum Grav. {\bf 21} (2004) R53-153, and references therein; M. Bojowald {\em et al.}, Phys. Rev. Lett. {\bf 95} (2005) 091302, and references therein.
\bibitem{vlhc} G. Ambrosio {\em et al.}, FNAL-TM-2149 (2001); W. Scandale and F. Zimmermann, Nucl. Phys. B Proc. Suppl. {\bf 177-178} (2008) 207; P. Limon, in eConf/C010107; G. Dugan and M. Syphers, CBN-99-15 (1999); A.D. Kovalenko, in
{\it Tsukuba 2001, High Energy Accelerators}, p2hc05; P. McIntyre,
in {\it Proc. Beyond 2010}, in press; and references therein.
\bibitem{wittn} See for example E. Witten, Phys. Lett. B{\bf 105} (1981) 267; Nucl. Phys. B{\bf 188}(1981) 513; M. Dine, W. Fishler, and M. Srednicki, {\it ibid.}{\bf 189}(1981) 575; S. Dimopoulos and S. Raby, Stanford ITP preprint (1981); S. Dimopoulos, S. Raby and F. Wilczek, Phys. Rev. D{\bf 24} (1981) 1681, and references therein.
\bibitem{gut1} See for example S. Raby, AIP Conf. Proc. {\bf 1078} (2009) 128;
J. Ellis, A. Mustafayev and K. A. Olive, 
arXiv:1003.3677, and references therein. 
\bibitem{kdgut}K. R. Dienes, E. Dudas and T. Gherghetta, Phys.Lett.B{\bf 436} (1998) 55; Nucl.Phys.B{\bf 537} (1999) 47; A. B. Kobakhidze, Phys. Lett. B{\bf 514} (2001) 131, and references therein. 
\bibitem{fr-min-all} H. Fritzsch and P. Minkowski, Phys. Lett. B{\bf 56} (1975) 69;
M. Gell-Mann, P. Ramond and R. Slansky, Rev. Mod. Phys. {\bf 50} (1978) 721;
P. Langacker, G. Segre and H. A. Weldon, Phys. Lett. B{\bf 73} (1978) 87; Phys. Rev. D{\bf 18} (1978) 552; P. Fayet, Phys. Lett. B{\bf 153} (1985) 397;
P.H. Frampton and S.L. Glashow, Phys. Lett. B {\bf 131} (1983) 340; {\it ibid.} {\bf 135} (1985) 515(E); 
R.N. Mohapatra, Phys. Rev. D{\bf 54} (1996) 5728;
Z. Berezhiani, I. Gogoladze and A. Kobakhidze, Phys. Lett. B{\bf 522} (2001) 107, and references therein.
\bibitem{neut1} See for example D. Wark, in {\it Proc. ICHEP02}, eds. S. Bentvelsen et al., (North-Holland,Amsterdam, 2003), Nucl. Phys. B (Proc. Suppl.) {\bf 117} (2003) 164; M. C. Gonzalez-Garcia, {\it ibid.}, {\bf 117} (2003) 186, and references therein.
\bibitem{gross} See for example G. G. Ross, {\it Grand Unified Theories}, (Benjamin-Cummings Publ. Co., Menlo Park, 1985), and references therein.
\bibitem{schwz1} See, for example, J. Schwarz, in {\it Proc. Berkeley Chew Jubilee, 1984}, eds. C. DeTar 
{\it et al.} (World Scientific, Singapore, 1985) p. 106, and references
therein.
\bibitem{gellmn} M. Gell-Mann, Physics {\bf 1} (1964) 63.
\bibitem{LFLi} See for example L.-F. Li, Phys. Rev. D{\bf 9} (1974) 1723 and rferences therein.
\bibitem{siggi} S. Bethke, Eur. Phys. J. C{\bf 64} (2009) 689.
\bibitem{pdg08} C. Amsler {\it et al.}, Phys. Lett. B{\bf 667} (2008) 1.
\bibitem{lewwg} S. Schael {\it et al.}, J. Abdallah {\it et al.}, M. Acciarri {\it et al.}, G. Abbiendi {\it et al.} and K. Abe {\it et al.}, Phys. Rept. {\bf 427} (2006) 257. 
\bibitem{gqw} H. Georgi, H. R. Quinn and S. Weinberg, Phys. Rev. Lett.{\bf 33} (1974) 451.
\end{thebibliography}
\end{document}